# METIS for ELT: optical alignment and testing of the IFU of the LM-band spectrometer

Lawrence Bissell[*a], Éamonn Harvey[a], Stephen Todd[a], Valentina Oyarzun[a], Alastair Macleod[a], Sandi Wilson[a], Alistair Glasse[a], Daniel Dicken[a], Felix Bettonvil[b,c], Bernhard Brandl[c]

[a]STFC - United Kingdom Astronomy Technology Centre (UK ATC), Edinburgh, UK;
[b]NOVA optical infrared instrumentation group at ASTRON, Oude Hoogeveensedijk 4, 7991 PD, Dwingeloo, The Netherlands
[c]Leiden Observatory, Leiden University, P.O. Box 9513, 2300 RA Leiden, The Netherlands

## ABSTRACT

The METIS LMS (LM-band Spectrometer) is a high-resolution (R>100,000) mid-infrared integral field spectrometer for the ESO ELT. The instrument provides the capability to measure instantaneous spectra across a 0.577 x 0.897 arcsecond field-of-view with a spatial sampling of 8.2-9.1 mas using an image slicing IFU. This paper presents the optical testing and alignment performed for the IFU sub-system. This includes the detailed characterisation of the image slicing mirror array measured via: a re-imaging test in which the slicer is positioned to re-image a point source to form 28 images due to the different tilts of the 28 slicer mirrors; and precise interferometric measurements of each of the image slicer mirrors. The paper will also give an overview of the optical alignment strategy employed during the IFU AIT process in order to reach the tight alignment requirements.

## 1 INTRODUCTION

The LMS (LM-band Spectrometer) is a METIS sub-system which provides the capability for high spectral resolution integral field spectroscopy at wavelengths between 2.7 and 5.3 microns (the atmospheric L and M bands). An image slicing integral field unit (IFU) provides spatially resolved spectra measured across a 0.577 x 0.897 arcsecond field of view. A block diagram of the optics of the LMS is shown in Figure 1.

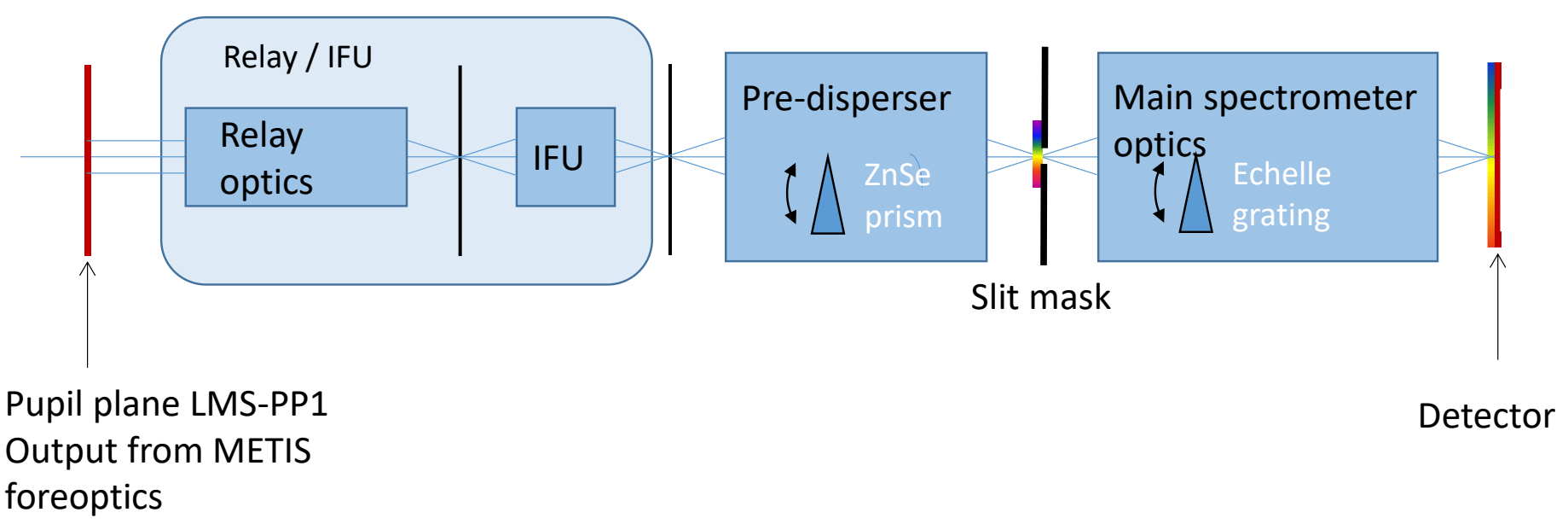

Figure 1. Block diagram overview of the optics of the LMS.

- Light is input to the LMS from the METIS Common Fore-Optics (CFO) at the input pupil plane (LMS-PP1). Relay optics then produce an image with a plate-scale of 0.02 arcseconds per mm.
- An all-reflective image slicing Integral Field Unit (IFU) with 28 slices then reformats the two dimensional field of view to produce the telecentric input slit for the pre-disperser.

[*]lawrence.bissell@stfc.ac.uk

- The pre-disperser is a low resolution prism spectrometer with an exit slit mask. This functions as a low spectral resolution monochromator, defining the bandpass within the LM band, acting as an order sorting filter for the main disperser.
- The main disperser is a high resolution spectrometer. A three mirror anastigmat (TMA) is used in double pass as the collimator and camera of the spectrometer, while a Germanium immersed echelle grating is place at the pupil to disperse the light. The instantaneous coverage of the spectrum is very small compared to the complete LM band so multiple exposures with rotation of the pre-disperser and main disperser are required to build up a continuous spectrum.
- In addition, a spectral slicer can be used to increase the instantaneous wavelength coverage by allowing multiple spectral orders to be observed simultaneously (similar to a cross-dispersing spectrometer). This bypasses the slit mask between the pre-disperser and main disperser, instead directing the low resolution spectrum from the pre-disperser onto a second slicing mirror which selects multiple pass-bands from within the spectrum. In this mode only a small number of slices from the IFU can be used so the spatial field of view is significantly reduced.

The IFU was designed such that the optical components could be manufactured using single point diamond machining to generate the multiple mirror facets on monolithic components, as previously applied on MIRI [1] and KMOS [2]. An overview of the optical layout is shown in Figure 2.

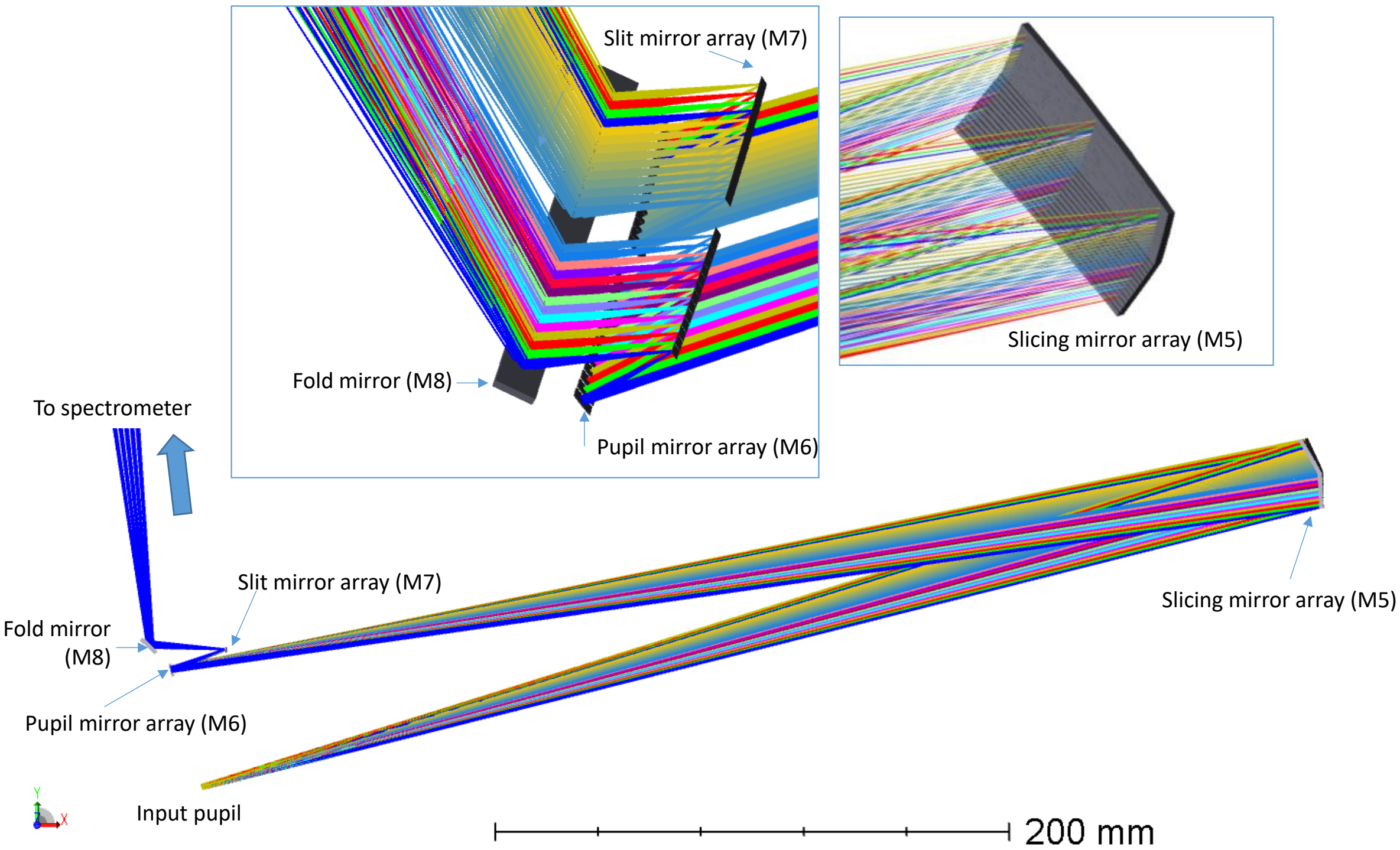

Figure 2. Optical layout of the IFU

Each facet of the image slicer (M5) is 1 mm wide and 50mm tall at room temperature, to give spatial sampling of 20.6 mas per slice. The surface of each mirror facet is spherical with a radius of curvature of 451.782mm at room temperature. The centres of curvature of the slices are all offset relative to one another along the long axis of the slice (equivalent to tilting each slice, since they are spherical) to produce a column of pupil images, one containing the light from a single slice.
The pupil mirror array (M6) is a linear array of spherical mirror facets, each 5 mm x 2.5 mm, placed at the pupil of each slice. The radius of curvature is around 42.7 mm concave but varies slightly along the array. In addition, the fold mirror (M8) is machined monolithically with M6 to simplify the alignment of the IFU.
The slit mirror array (M7) is another linear array of spherical mirrors, this time each 2mm x 2.48mm and with a radius of curvature of around 44.2mm. Both M6 and M7 are stepped along the optical axis in order to form a slit image on a curved focal plane with radius of curvature 1000mm (centre of curvature away from the IFU) to improve the performance of the pre-disperser.

The three optical components have now been manufactured and delivered, as shown in Figure 3. The image slicer was manufactured by Canon, while the other IFU optical components were manufactured by Durham University Centre for Advanced Instrumentation.

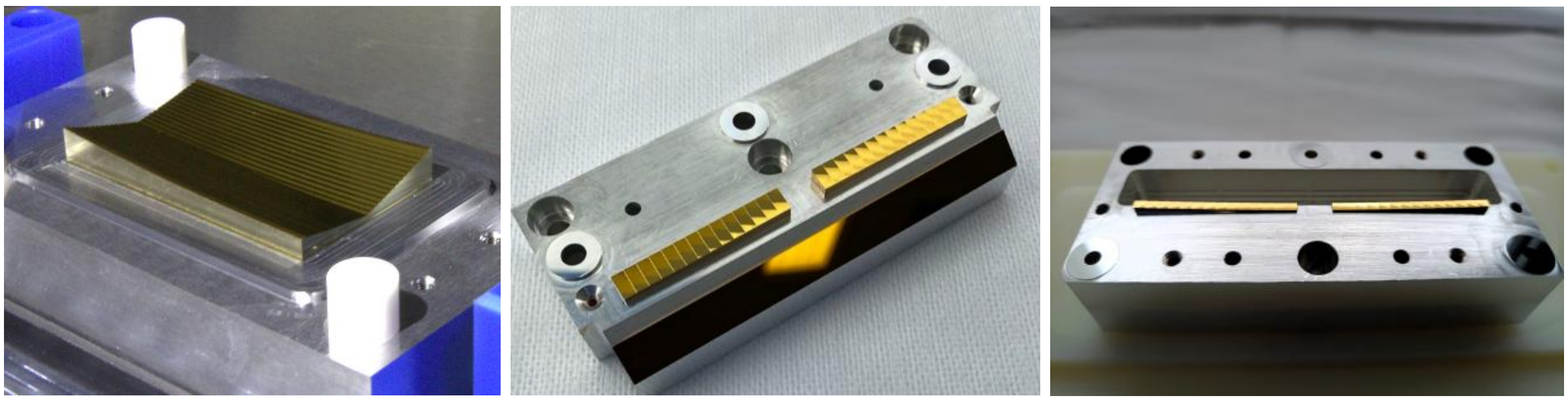

Figure 3. Images of the IFU optical components: (left) image slicer, (centre) pupil mirror array and fold mirror, (right) slit mirror array.

The verification of three of the IFU optical manufacturing requirements regarding the image slicer was performed at the UK Astronomy Technology Centre in the METIS LMS assembly facility. These requirements are shown in Table 1.

Table 1. List of IFU requirements under test for the image slicer.

| | Requirement |
|---|---|
| **IFU REQ 10.** | The relative positions of the centres of curvature must be maintained to within the following tolerances: Cx ±10 µm; Cy ±10 µm; Cz ±50 µm. |
| **IFU REQ 13.** | The maximum slope error relative to the nominal surface must be less than 25 microrad to maintain pupil image quality at the pupil mirror array. |
| **IFU REQ 14.** | RMS surface form area in any 1mm x 5mm sub-aperture to be < 10nm. |

The following sections detail the optical testing performed to verify these requirements and the planned optical alignment of the IFU sub-assembly. Firstly, a re-imaging performance test setup and key results will be described. Then an interferometric test will be detailed. Finally, an overview of the IFU optical alignment strategy that will be employed during the Alignment Integration and Testing (AIT) process will be presented.

## 2 REIMAGING PERFORMANCE TESTING

To test the position of the centres of curvature of each of the mirror slices (IFU REQ 10.), the slicer is used to reimage a point spread function (PSF) produced by the µPhase 2HR interferometer laser into 28 different images near the centre of curvature of each of the slices. The test was designed to provide an upper bound on the centre of curvature position to allow for acceptance of the slicer.

### 2.1 Test Setup

The test setup is shown in Figure 4. Light is focused by the interferometer using an F/5 converging lens, 255mm from the output. It then passes through a central hole in the alignment screen before reflecting off the 28 image slicing mirrors, positioned at a distance equal to the radius of curvature of each of the slices (451.782mm). The light then comes to focus at the alignment screen approximately at the centre of curvature of each slice (the precise position is determined from Zemax modelling).

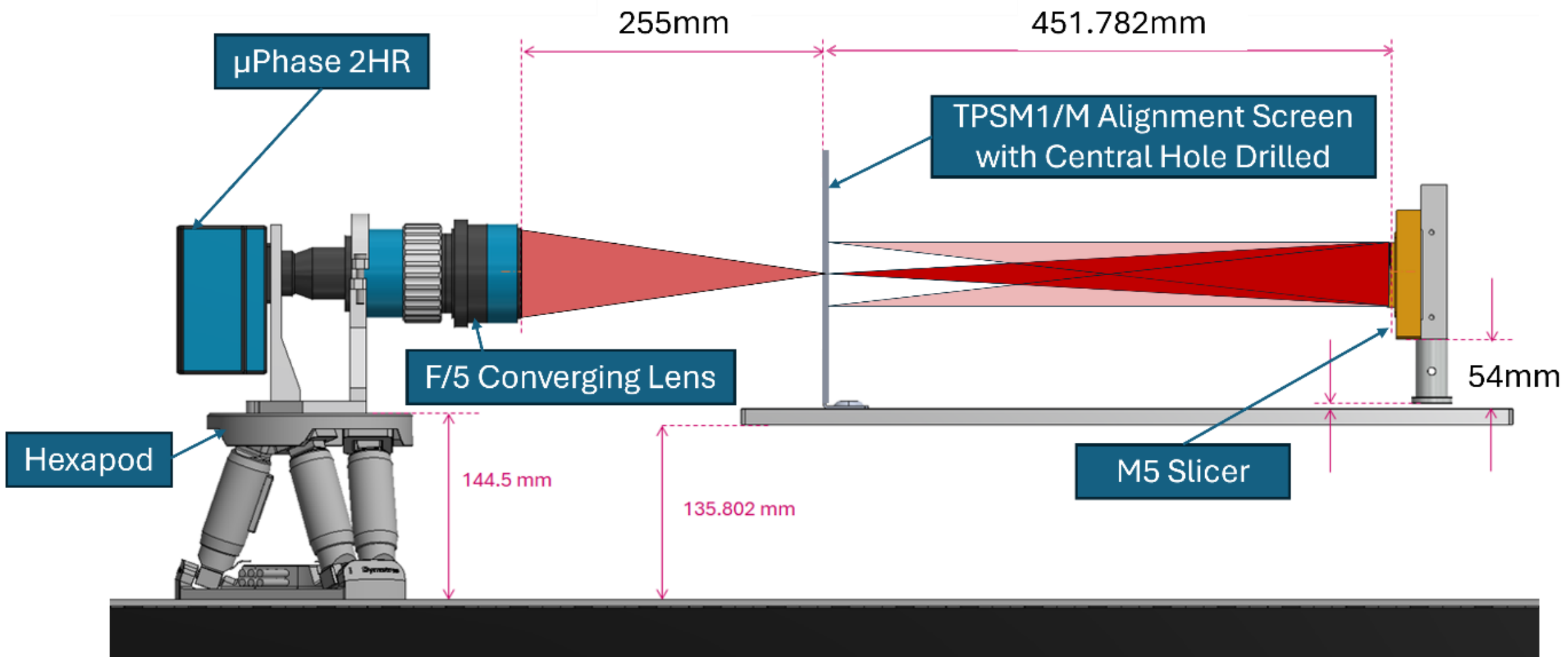

Figure 4. Diagram of reimaging performance test setup. Only the light reflected from 2 of the slices is shown for clarity – rather than the full 28 images that form on the alignment screen.

## 2.2 Alignment Procedure

To ensure good alignment between the interferometer focus and the image slicer, the interferometer was first offset to align the interferometer focus to the centre of curvature of one of the central slices (with the alignment screen removed). This configuration allows the direct interferometric measurement of this slice. The position of the interferometer was optimised by shifting laterally to minimise focus, tip, and tilt of the resultant interferogram. The interferometer was then shifted by 2.63mm vertically upwards and 2.61mm away from the slicer to position the interferometer focus at the correct position at the centre of the slicer.
Once this was complete, the alignment screen was placed in the optical path and shifted vertically and in focus to align the hole with the interferometer focus. The test setup after alignment is shown in Figure 5.

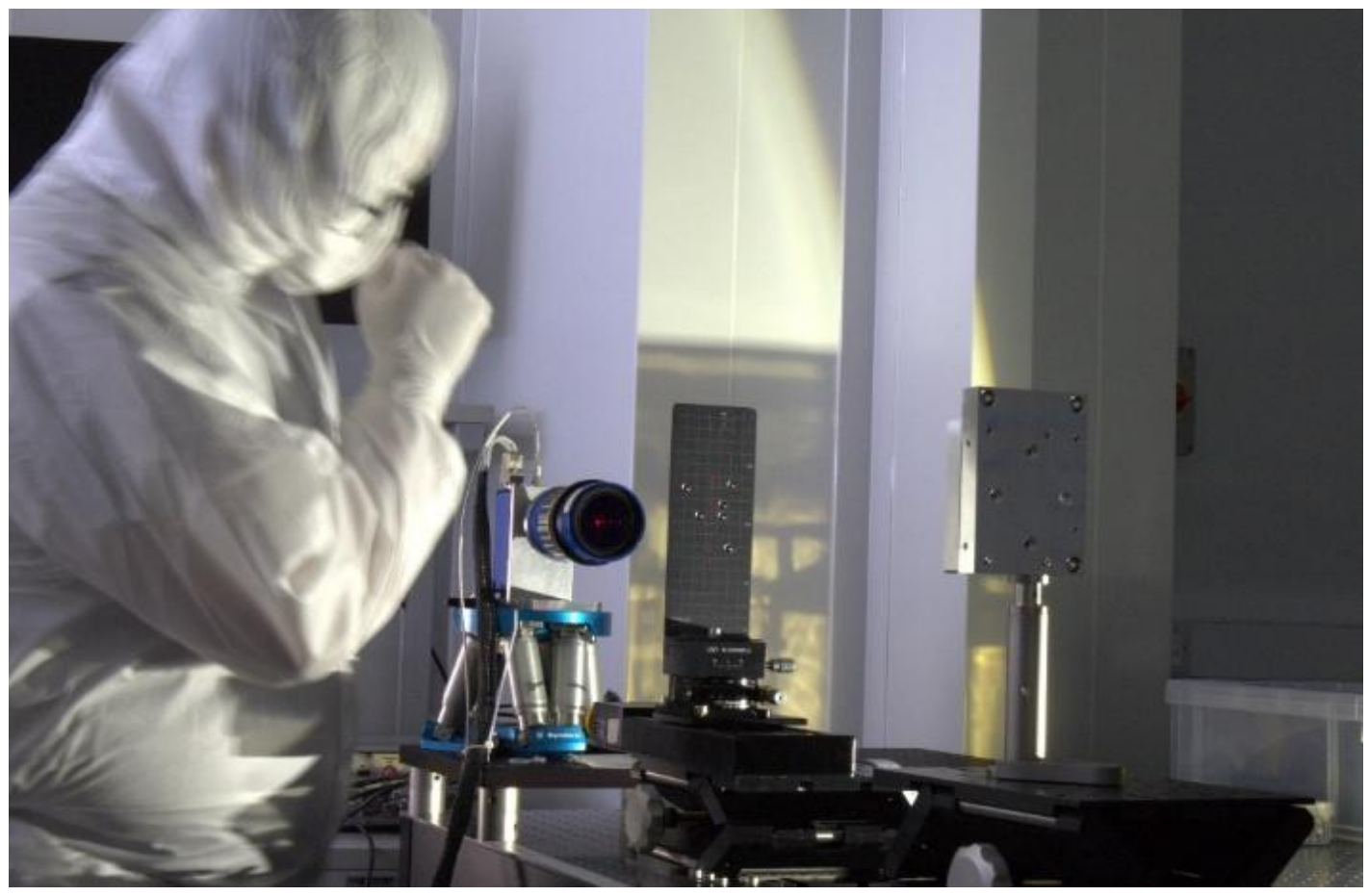
Figure 5. Image of optical test setup after alignment.

## 2.3 Data Acquisition

Following the successful set up and alignment of the reimaged PSFs on the target, the PSFs were imaged using a DSLR camera. This was done over the course of two days, as it was realised after the first day that additional angles and images would be required to get a better statistical sampling of the data during analysis. The first data set consisted of 8 images for which the lighting, SNR, angles, resolution, visibility of the gridlines on the target were of sufficient quality, see Figure 6. It was noted that the PSF that appeared most out of specification was on a grid line during these measurements.

On the second day several more images were taken to increase the data set. Additionally, images were taken with the target tilted so that PSFs that previously lay on a grid line were repositioned. In total a sample of ~50 appropriate images were used for analyses.

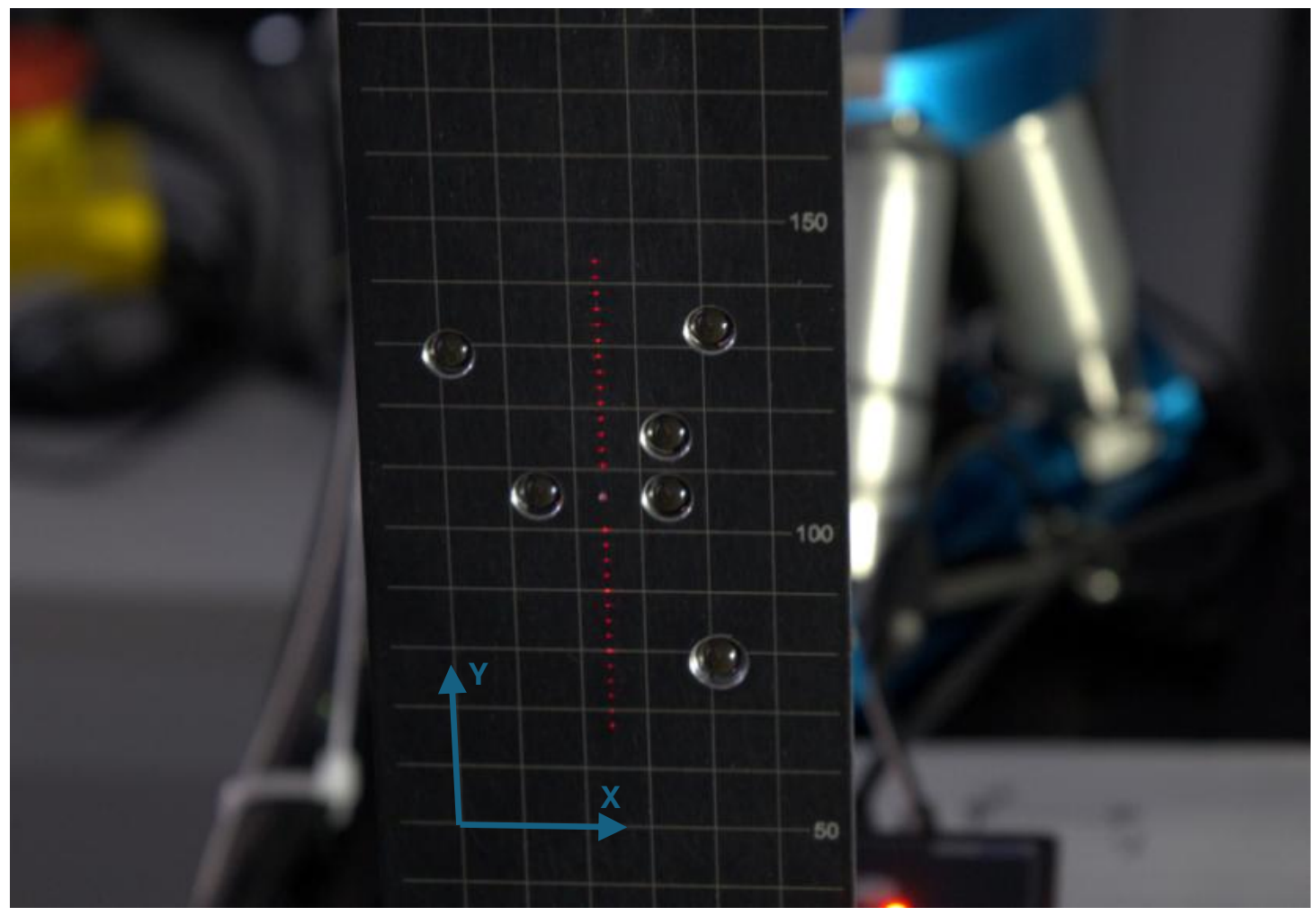

Figure 6. Photograph of the PSFs reimaged from the M5 image slicer onto the alignment screen.

## 2.4 PSF Fitting

To prepare the images to perform PSF fitting they initially needed to be cropped, unwarped and scaled. A script was written in Python to do this. The script performs the following steps:

1. A pixel-to-mm scale is computed from two known reference points.
2. A perspective transform is built and the image unwarped.
3. The image is converted to grayscale (extracting the red layer for the PSFs) and smoothed.
4. A crop region around the image centre is defined to exclude bright spots in the background from being spuriously identified as PSFs.

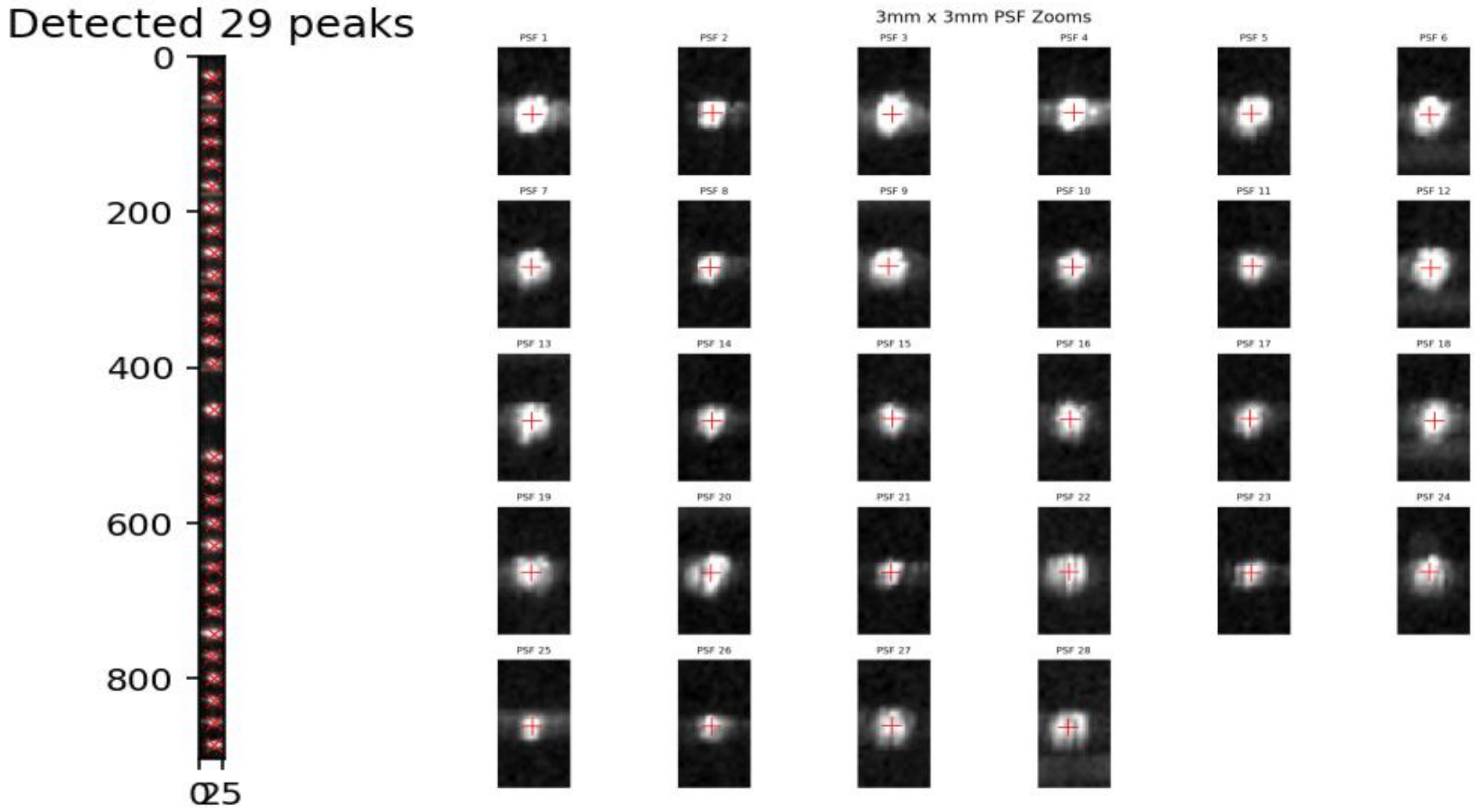

Figure 7. (left) A cropped series of PSFs with the 29 brightest spots detected. The central bright spot is the pinhole for which the powered interferometer beam comes through, which is removed from further analysis. (right) 28 subplots showing zoomed images of 28 identified PSFs that correspond to reimaged slices from the M5 slicer. Plots like these were done for each individual image and analysed as a quality check step.

Once this has been done, a Python script is used to perform the PSF fitting:

1. Candidate PSFs are identified with a two-stage detection pipeline (see Figure 7):
    a. local peak detection and;
    b. clustering to remove duplicate detections.
2. For each candidate PSF a Moffat model is nonlinearly fitted.
3. The fit results are sanity-checked (width/amplitude limits, realistic wings) and stored.
4. The verified peak centroids are then converted from pixels-mm via the previously identified scaling.

For each image, a simple geometric validation is then performed by fitting a line to the PSF positions and calculating the residuals (RMS and peak-to-valley). One example is shown in Figure 8.

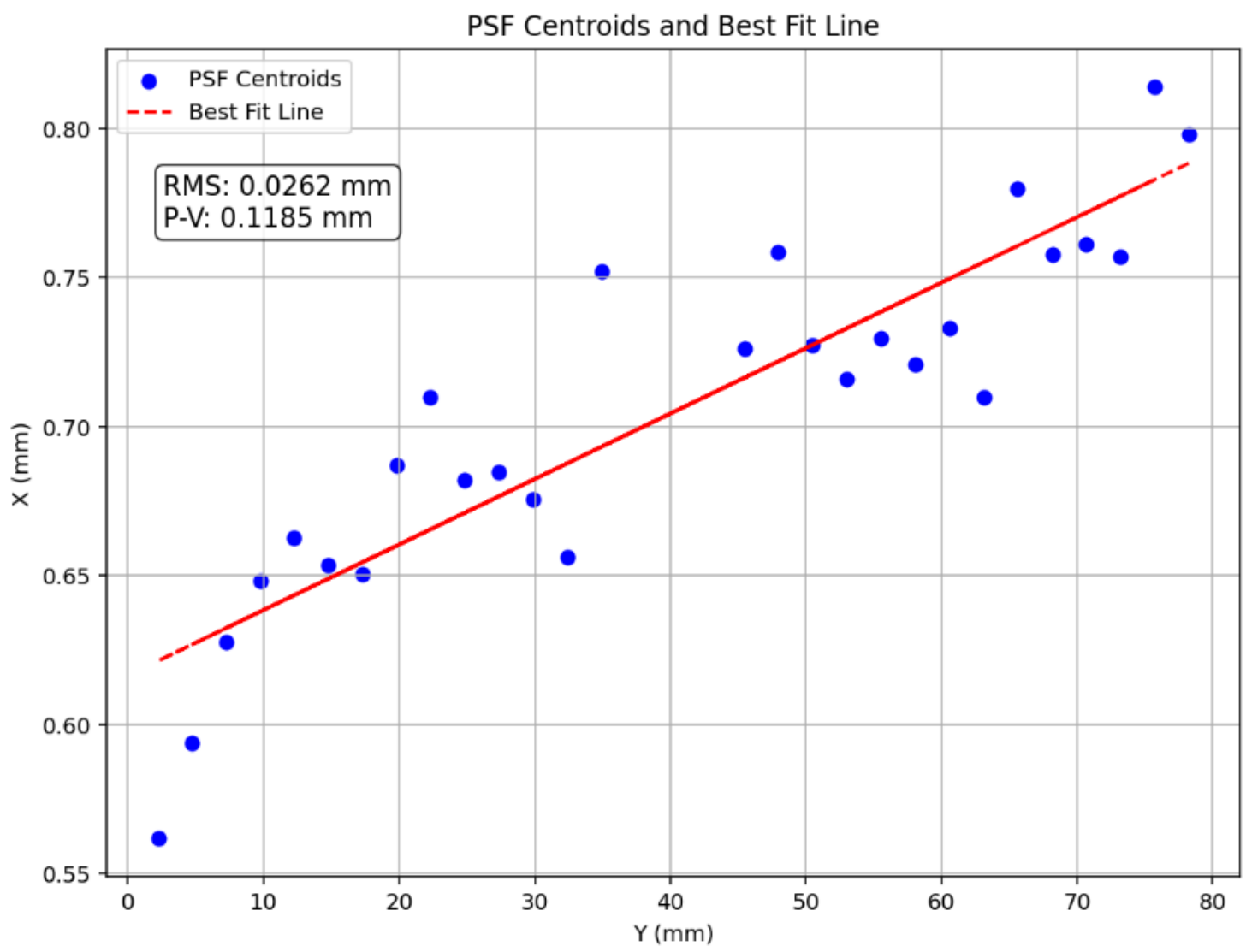

Figure 8. Example PSF centroiding from Moffat fits of measurements from an analysed image.

Once the PSFs have been fitted for each image, the resulting measurements are combined:

1. Per-image transforms and scale factors are applied to the fitted centroids so all data share consistent units and indexing.
2. The scaled measurements are concatenated into a single table.
3. Per-measurement centroid uncertainty is assigned (roughly tied to SNR).
4. Weighted means and uncertainties for each slice PSF position across the image stack is computed.
5. A linear fit is performed to the extracted PSF positions to remove bulk tilt of the alignment screen followed by the subtraction of the nominal X and Y position of the PSFs as modelled in Zemax.

### 2.5 Results

The requirements for the re-imaging PSF position is shown in
Table 2. This is a factor of two larger than the centre of curvature accuracy requirement due to the test setup.

Table 2. Breakdown of IFU REQ 10, showing the required centre of curvature and re-imaging PSF position accuracy requirements for each direction. Cx corresponds to the across slice direction, Cy the along slice direction, and Cz the focus direction.

| | Requirement | Centre of Curvature Accuracy Requirement | | Re-imaging PSF Position Accuracy Requirement | |
|---|---|---|---|---|---|
| **IFU REQ 10.** | The relative positions of the centres of curvature must be maintained to within the following tolerances: | Cx | ±10μm | Cx | ±20μm |
| | | Cy | ±10μm | Cy | ±20μm |
| | | Cz | ±50μm | Cz | Not tested. |

The Cx and Cy residual plots are shown in Figure 9. These show that the accuracy requirement does not appear to be met for several of the 28 slices in the both the X and Y directions:

- 13 of the 28 slice measurements are outside of specified requirement in both X and Y, although there is a significant error bars in the measurement.
- 5 of the slice measurements are out of specification in X within measurement error.
- 3 of the slice measurements are out of specification in Y within measurement error.

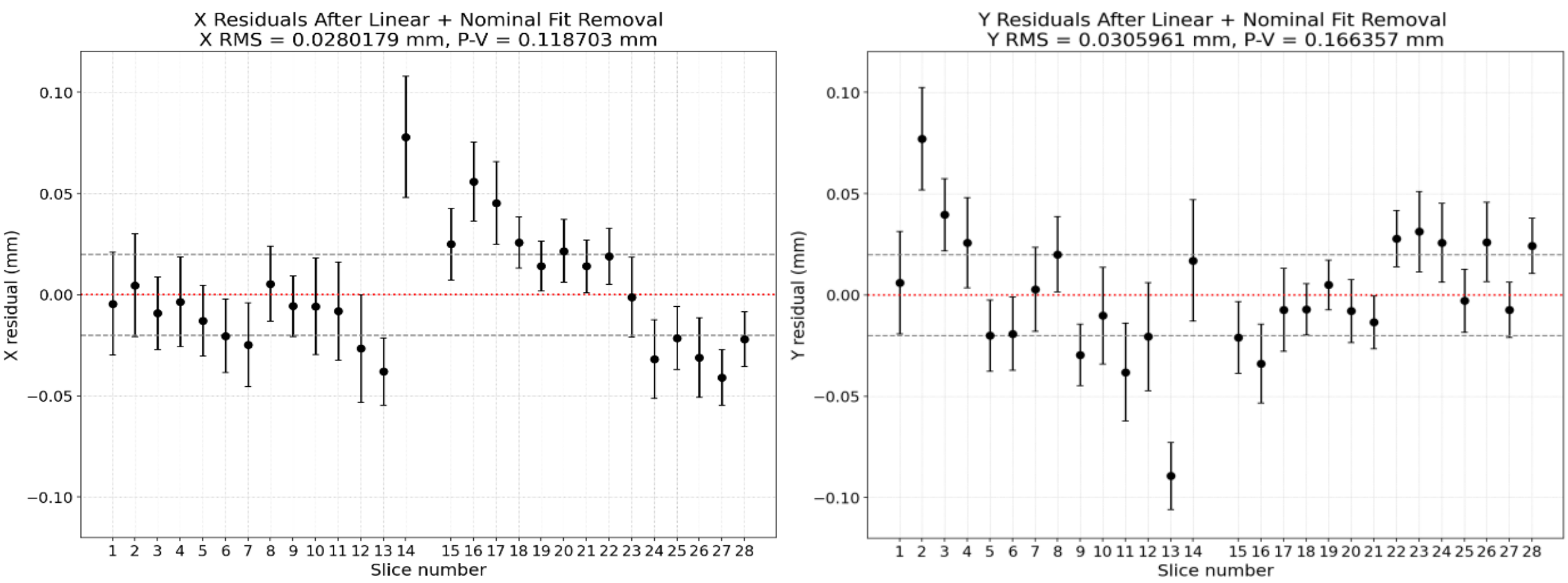

Figure 9. (left) Final X residuals after linear and nominal fit removal plotted against the slice number with a gap indicating the centre of the slicer. (right) Final Y residuals after linear and nominal fit removal plotted against the slice number with a gap indicating the centre of the slicer.

The centre of curvature variance between the slices was then calculated from the peak-to-valley (PV) of the measurement residuals – shown in Table 3.

Table 3. Comparison of Cx and Cy centre of curvature accuracy requirement with the measured variance.

| | Centre of Curvature Accuracy Requirement | Extracted Centre of Curvature Variance |
|---|---|---|
| **Cx** | ±10μm | **±29.7μm** |
| **Cy** | ±10μm | **±41.6μm** |

### 2.6 Conclusion

This test indicated that there is significant variation in the centre of curvature position of the slices in both X and Y – outside of the specification of the slicer. In addition, there is a large shift in the centre of curvature position between slices 13 and 14 with slices 1-13 and 14-28 group much more closely together. This indicated that the linear fit may have artificially tilted the output to minimise the RMS – however, there was no way to check this in the initial setup.
This required further confirmation with interferometric testing to give a more complete description of the slicer performance.

## 3 INTERFEROMETRIC TESTING

The interferometric test takes an interferometric measurement of each of the slices to allow the verification of IFU REQ 13. and IFU REQ 14. In addition, the position of the slicer relative to the interferometer is recorded to measure the relative position of the centres of curvature of each of the slices (IFU REQ 10.).

### 3.1 Test Setup

The test setup is shown in Figure 10. Light is focused by the interferometer using an F/5 converging lens, 255mm from the output. The image slicer is then positioned such the centre of curvature of one of the slices lies at the focus of the interferometer. This is done by shifting the linear stage to nominally align the slicer along the Y axis before fine alignment in X, Y, and Z is performed using the hexapod by minimising focus, tip, and tilt of the measured interferogram. The position of the hexapod and linear stage are then recorded as well as the measured interferogram of the slice. This process is repeated for each of the 28 slices.

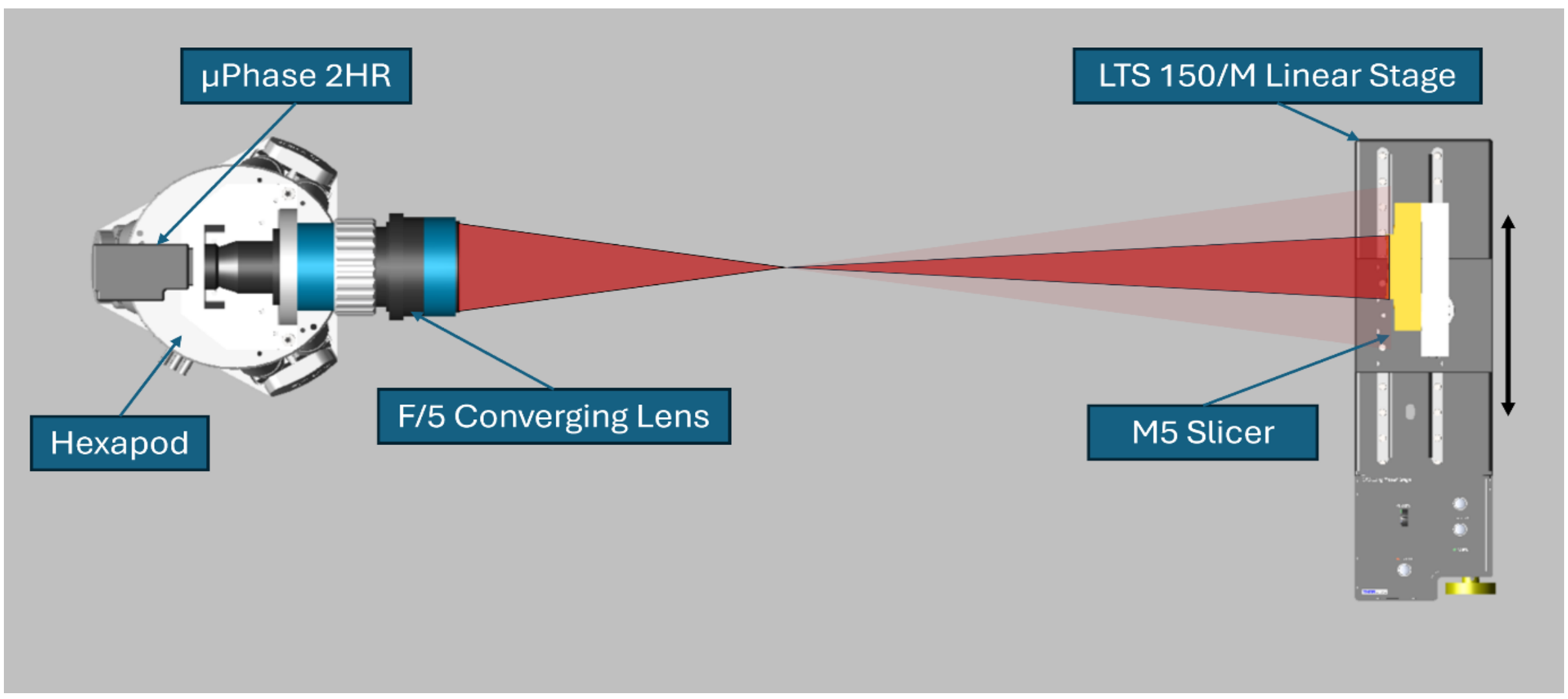

Figure 10. Diagram of interferometric test setup. The image slicer is tilted 90 degrees with respect to the previous test such that the slices are aligned with the linear stage axis and the centres of curvature lie in a horizontal plane. F/5 beam overfills slicer aperture.

### 3.2 Alignment Procedure

To ensure good alignment between the hexapod Y axis and the linear stage axis, the test was setup as shown above with the centre of curvature of slice 1 at the interferometer focus. Then both the hexapod and linear stage were moved 10mm before the slice 1 was realigned in X, Y, and Z using the hexapod by minimising focus, tip, and tilt of the interferogram. This required a 0.094mm focus shift (Cz) – corresponding to a tilt between the two stages of 0.5737°. The hexapod was then tilted by this angle and a new coordinate system setup to align the Y axis with the axis of the linear stage.

Once this calibration had been performed, the process was repeated again for slice 1 verifying the coalignment of the two stages and measuring the small calibration residuals:

- Cx: 6µm out-of-plane residual over 10mm Y travel.
- Cy: 10µm in-plane residual over 10mm Y travel.

These were assumed to be linear and were used to calibrate the linear stage positions.

### 3.3 Data Acquisition

Measurements were taken on a floated optical bench in the METIS ISO 6 cleanroom over the course of 1.5 hours with the lab at 20°C. For each slice the linear stage was shifted to nominally align the slicer along the Y axis and a corresponding nominal shift was applied along the Z axis of the hexapod to coarsely align the slice to the interferometer and a measurement was then taken. The interferometer then reported residual defocus, tip, and tilt which was used to iteratively re-position the interferometer until the residual was <10nm in focus and <1" in tip/tilt. The following was then recorded:

- X, Y, Z coordinates of hexapod.
- Y position of linear stage.
- Defocus, tip, and tilt residuals.
- The final measured interferogram of the slice.

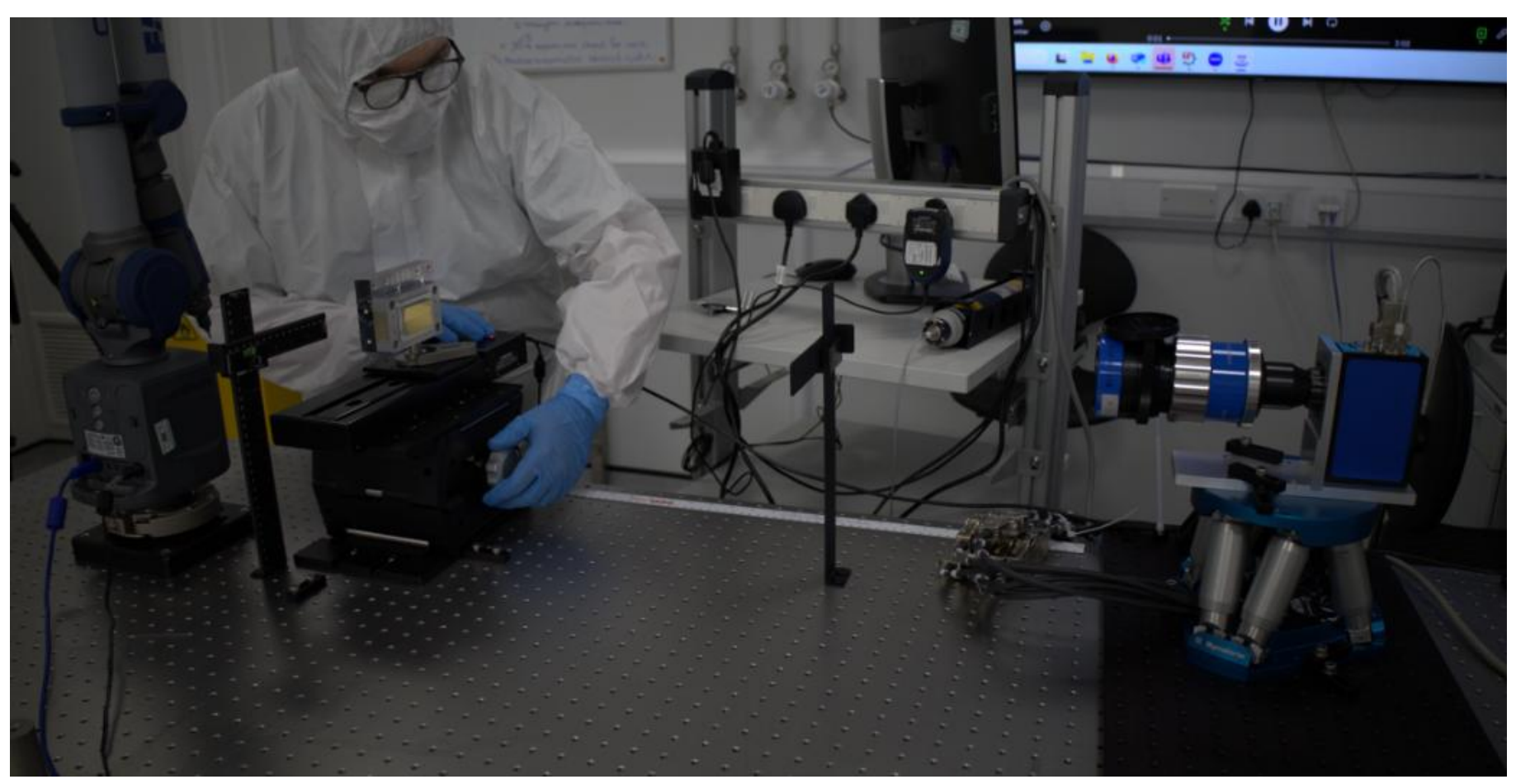

Figure 11. Image of optical test setup during alignment.

### 3.4 Centre of Curvature Analysis

To calculate the centre of curvature of each of the slices, the following steps were performed:

1. The recorded linear stage position for each slice was calibrated using the residuals from the alignment procedure.
2. The calibrated linear stage shifts were then added to the hexapod coordinates to derive the positions of the centre of curvature in the hexapod frame of reference.

Since the slicer was not perfectly aligned with the hexapod reference frame, a coordinate transformation was applied to the results in order to find the centres of curvature in the slicer reference frame:

1. A -0.83° rotation was applied about the X axis which removed all tilt residual from the Cz measurements.
2. A 0.067° rotation matrix about the Z axis was applied to correct for the clocking of the interferograms as seen on the interferometer camera.

The nominal Cx, Cy, and Cz positions were then subtracted from the data before the average value of each was subtracted to find a zero-mean residual.

The resulting Cx, Cy, and Cz residual plots are shown in Figure 12.

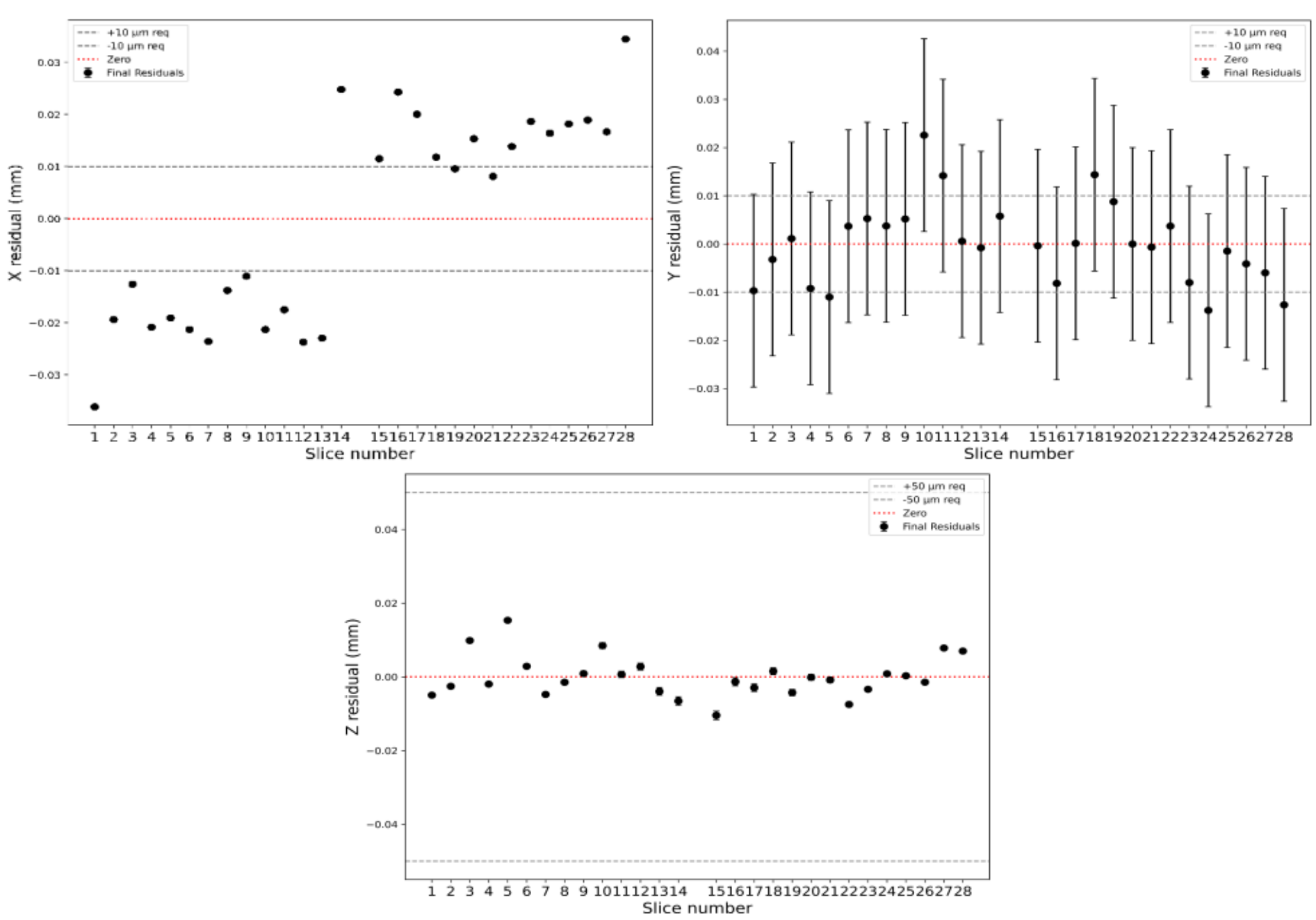

Figure 12: (top-left) Final X residuals plotted against the slice number with a gap indicating the centre of the slicer. (top-right) Final Y residuals plotted against the slice number with a gap indicating the centre of the slicer. (bottom) Final Z residuals plotted against the slice number with a gap indicating the centre of the slicer.

These results confirm the variation in centre of curvature seen in the previous test (summarised in Table 4). The results for Cx show a step of ~50µm between slices 13 and 14 which results in the performance being out-of-specification. The Cy performance was shown to be better than previously estimated, however remains slightly out-of-specification but within experimental error of the test, while the Cz performance is within specification.

Table 4. Comparison of Cx and Cy centre of curvature accuracy requirement with the measured variance.

| **Requirement** | | | **Measured Values** |
|---|---|---|---|
| **IFU REQ 10.** | Cx | ±10µm | **±36.1µm** |
| | Cy | ±10µm | **±22.6µm** |
| | Cz | ±50µm | **±15.3µm** |

### 3.5 Slope Error Analysis

To assess IFU REQ. 13 the interferogram of each slice was analysed individually. The image of each interferogram is shown overlaying the interferometer field-of-view in Figure 13.

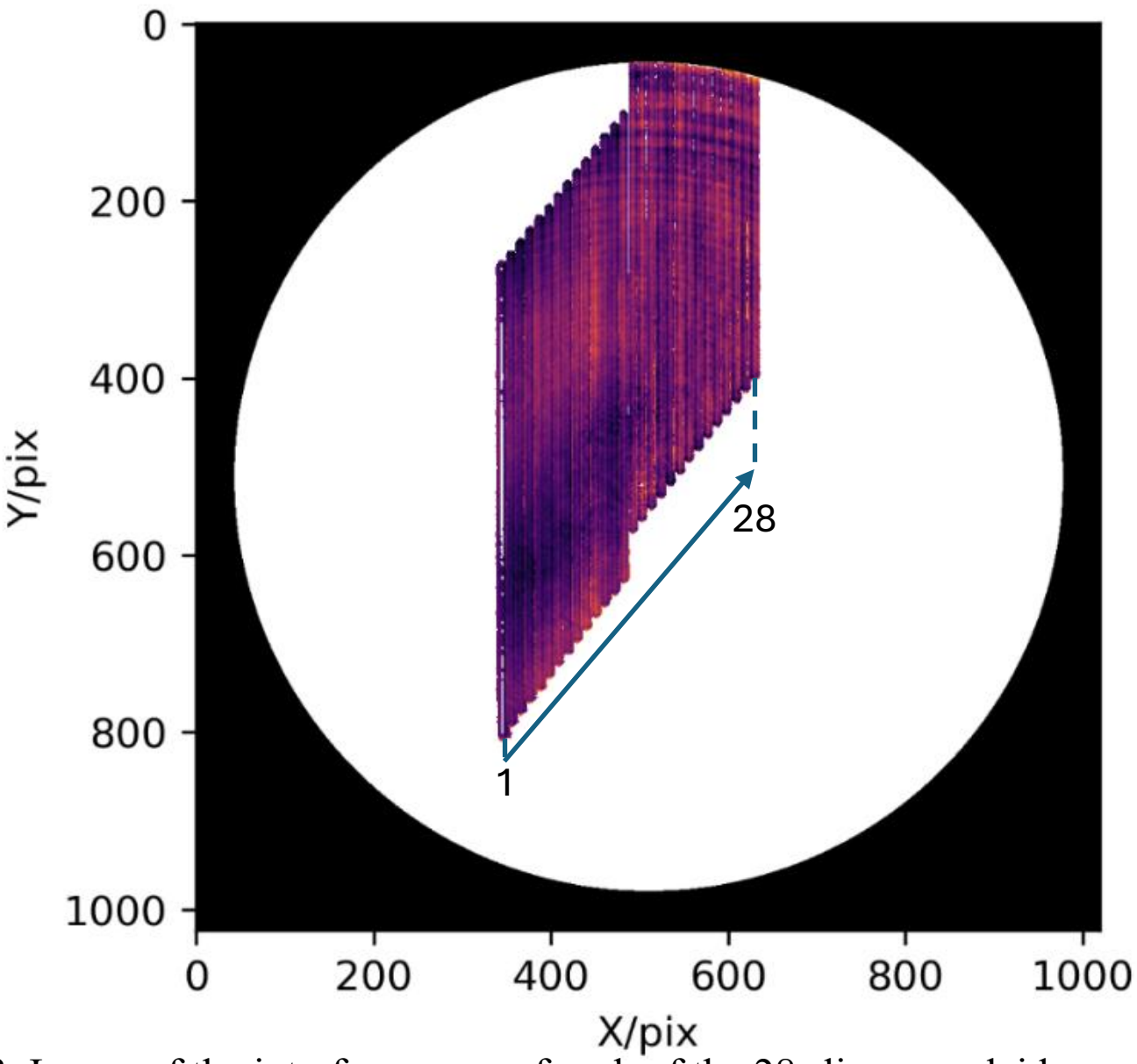

Figure 13. Image of the interferograms of each of the 28 slices overlaid on an aperture matching the interferometer field-of-view. The slice number increases from left to right across the image. Slices 15-28 are vignetted by the edge of the field-of-view.

To analyse the slope error of the slices, the interferograms needed to be processed using a Python script. The script performs the following steps:

1. The data file for each interferogram is read into Python using the *prysm* [3] package to form an Interferogram object.
2. The interferogram is then cropped to a rectangle bounding the region with valid data.
3. Residual piston, tip and tilt is then subtracted from the measurement.
4. The mean of the data is taken in the across-slice direction leaving a 1D profile along the length of each slice. These are plotted for slices 1-14 and 15-28 in Figure 14 (A and B).
5. The 1D profile is then cropped to the 50mm length of each slice to remove edge effects largely due to image blur (shown in Figure 14 C and D).

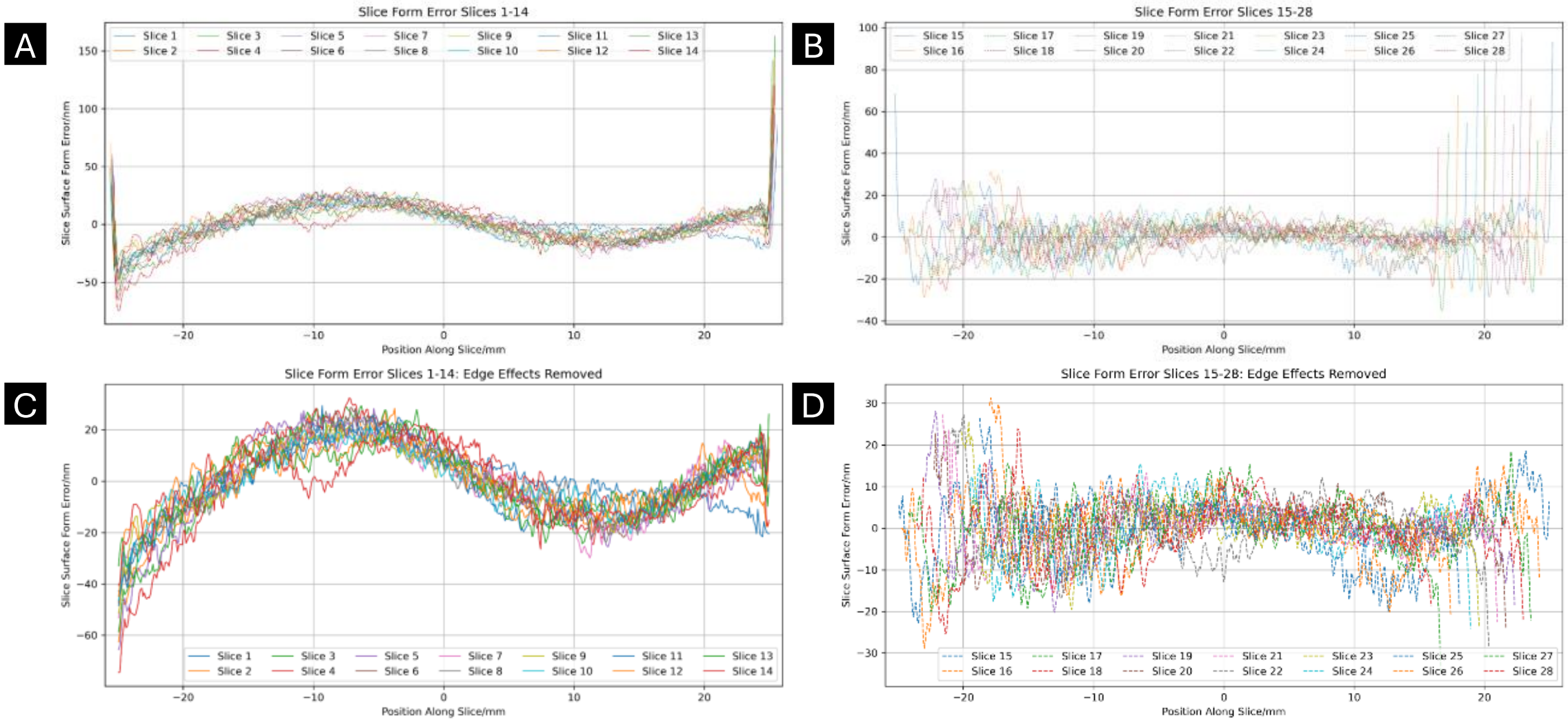

Figure 14. Uncropped slice surface form error measurements for slices 1-14 (A) and 15-28 (B). Cropped slice surface form error measurements for slices 1-14 (C) and 15-28 (D) to remove edge effects due to interferometer focus and diffraction.

6. For each position along the slice a 1mm window is then defined and a line is fit to the windowed data to calculate the slope error. The result of this is shown in Figure 15.

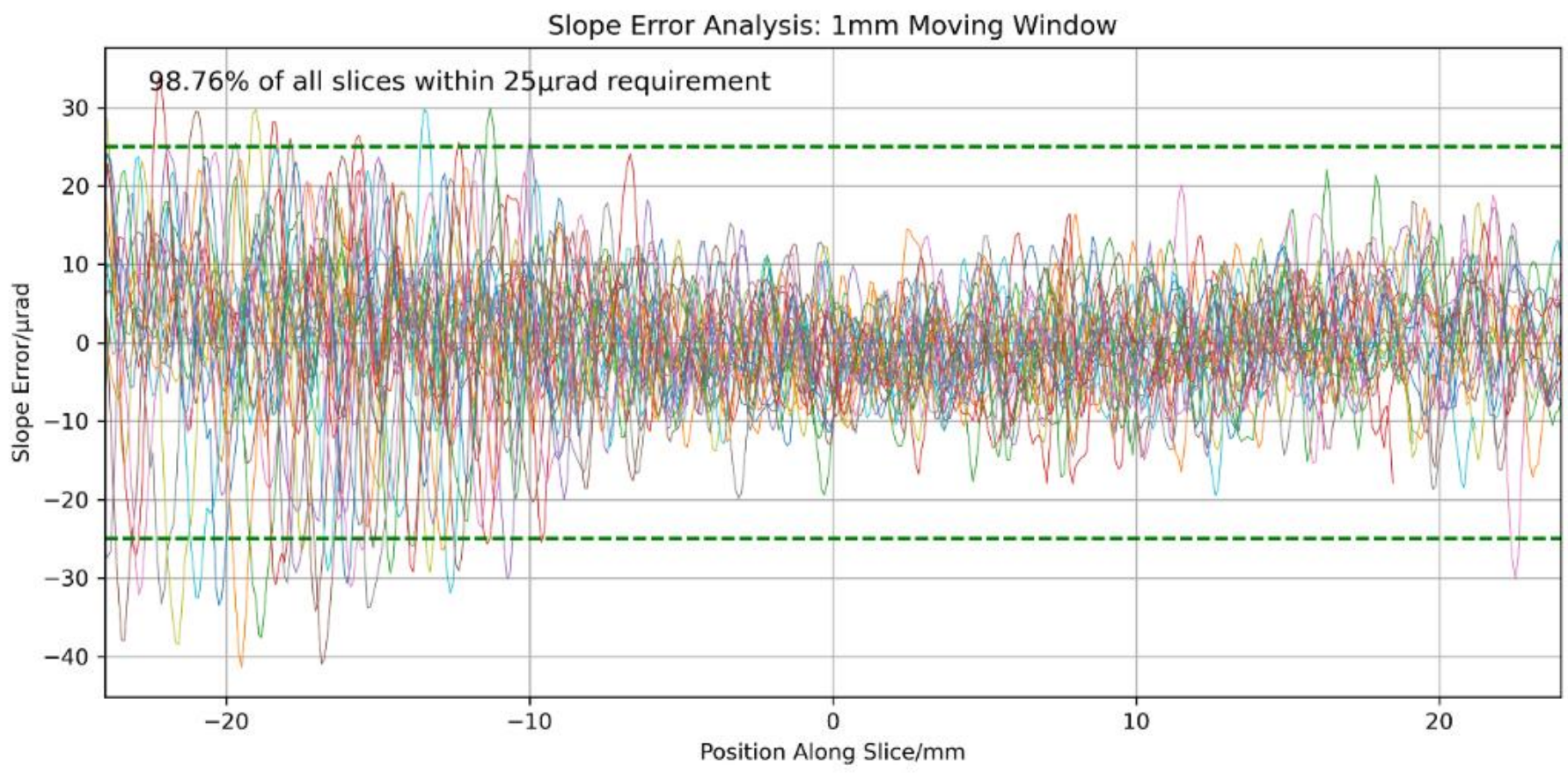

Figure 15. Slope error of each of the 28 slices. The green dashed lines indicate the 25µrad requirement.

This result shows that **98.76%** of the total length of all of the slices meets the 25µrad requirement. The RMS slope error is **7.99µrad**, while the maximum slope error is **41.43µrad**.
Looking more closely at the data we see that the region in which most of the non-compliance is measured is near the edge of the interferometer field-of-view (see Figure 16). In this region, there can be seen a significant ripple due to the residual of the interferometer across each of the slices, which accounts for this non-compliance. Once this data is removed, **99.90%** of the total length of all of the slices meets the 25µrad requirement.

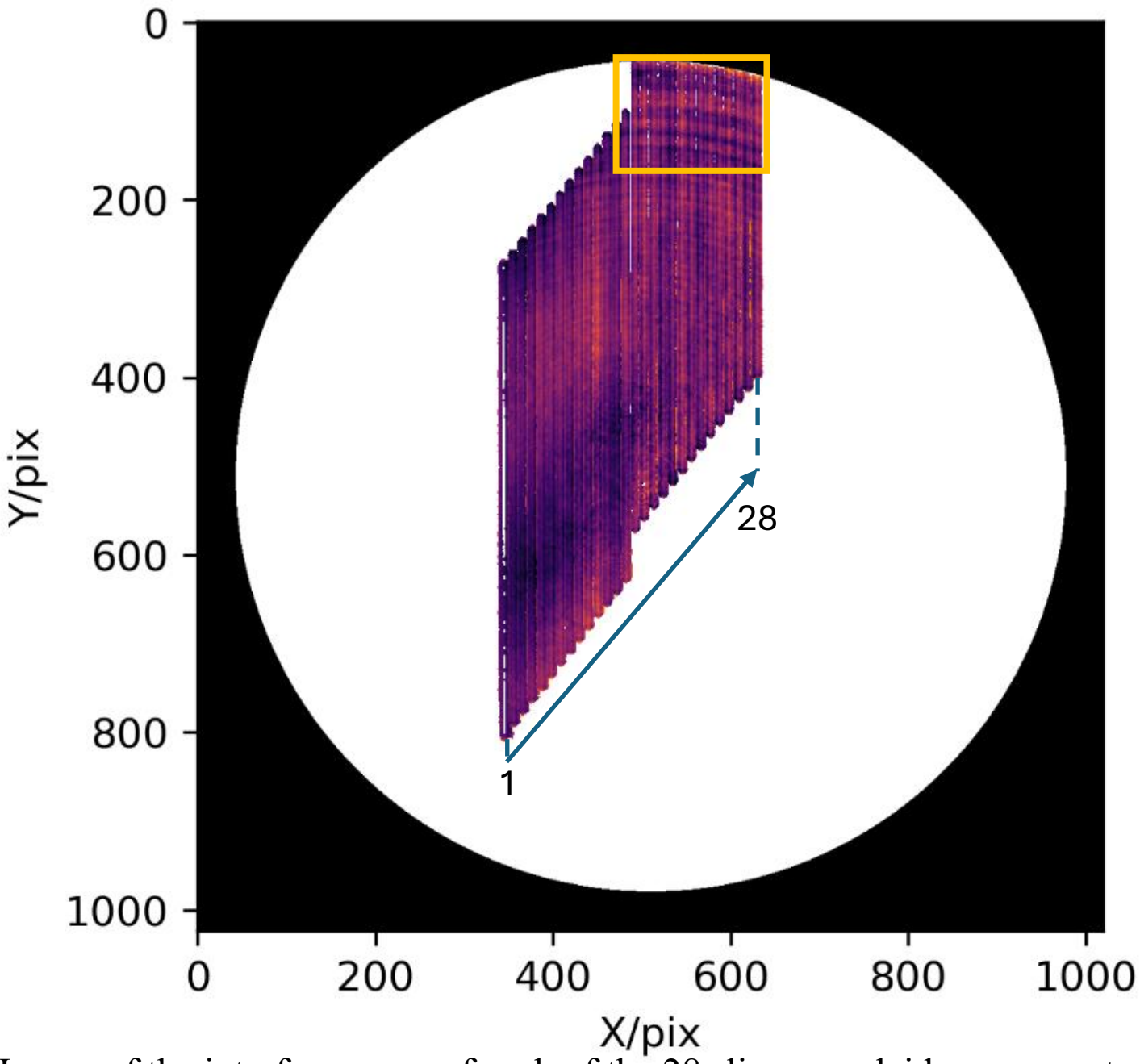

Figure 16. Image of the interferograms of each of the 28 slices overlaid on an aperture matching the interferometer field-of-view. The yellow box highlights the region in which most of the non-compliant slope error is measured.

With these measurements and data provided by Canon on manufactured dummy slices, IFU REQ 13 is deemed to be compliant.

Table 5. Comparison of maximum slope error requirement with the measured values.

| IFU REQ 13. | | Measured Values |
|---|---|---|
| **Maximum Slope Error** | <25µrad | **7.99µrad RMS** |
| | | **99.90% within specification after removal of interferometer artifacts** |
| | | **PVq/2 = 23.7µrad for q = 99.9%** |

### 3.6 Surface Form Error

To assess IFU REQ. 14, the interferogram of each slice was again analysed individually, with interferograms processed using a Python script. The script performs the following steps:

1. The data file for each interferogram is read into Python using the Prysm package to form an Interferogram object.
2. The interferogram is then cropped to a rectangle bounding the region with valid data.
3. Residual piston, tip and tilt is then subtracted from the measurement.
4. A 5mm x 1mm rectangular mask is then defined and the slice measurement is masked every 2.5mm down the length of the slice.
5. Residual piston, tip and tilt is subtracted from each of the masked measurements.
6. The mean of the masked data is taken in the across-slice direction leaving a 1D profile in the along-slice direction, to account for aperture diffraction and blurring effects seen in the interferograms.
7. The RMS surface form error of each 1D profile is recorded.

The resulting distribution of form errors is shown in Figure 17. This shows that **98.85%** of all sub-apertures are within the 10nm RMS requirement. The mean RMS form error is **4.57nm**, while the maximum RMS form error is **11.67nm**. This is slightly above the 10nm budget but will have an insignificant effect on the total LMS wavefront error.

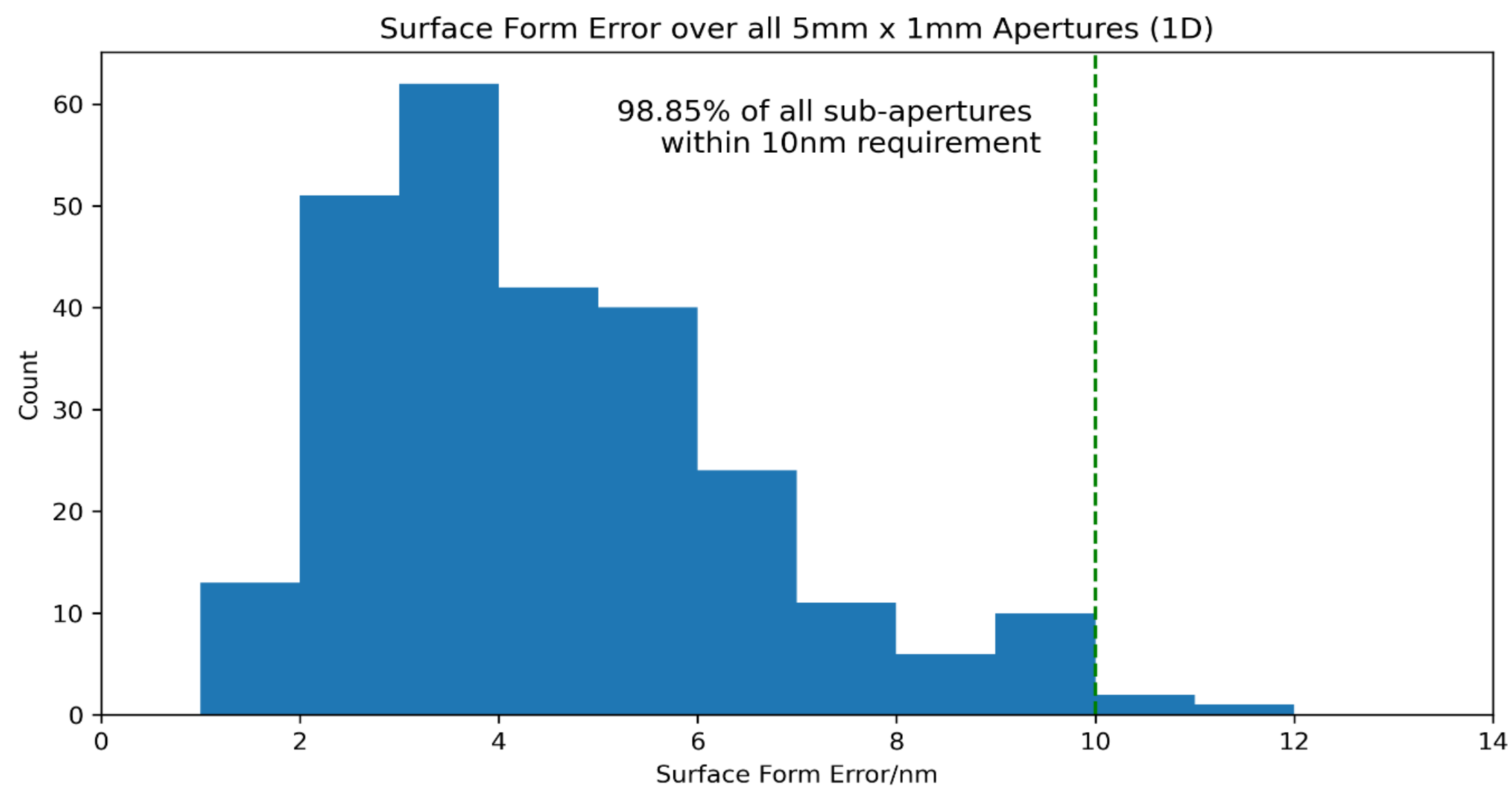

Figure 17. Surface form error distribution of each of the 5mm x 1mm apertures of all 28 slice 1D profiles.

Table 6. Comparison of maximum surface form error requirement with the measured values.

| **IFU REQ 14.** | | **Measured Values** |
|---|---|---|
| **Surface form error over 5mm x 1mm sub-apertures** | <10nm RMS | **Mean: 4.57nm RMS** |
| | | **98.85% of all sub-apertures within 10nm requirement** |
| | | **Max: 11.67nm RMS** |

## 4 IFU SUB-ASSEMBLY AIT PLAN

The IFU sub-assembly contains the image slicer (M5), the pupil mirror array (M6), the slit mirror array (M7), and the fold mirror (M8) machined monolithically with M6 as detailed in section 1. M6-8 are mounted together on a precision manufactured mount to form the slit-pupil group (SPG), which can be aligned to M5 as a single unit – simplifying the overall IFU alignment. The measured manufacturing error of the image slicer imposes a tight alignment accuracy requirement on the IFU mirrors.

To achieve this, the following tests will be performed:

- **Test 1:** Light from the interferometer (INTF) is focused onto the slit mirror array. The light passes through the IFU in double pass, reflecting off tooling balls acting as return spheres (BGA04) placed at the image slicer focal plane. The SPG is adjusted to align and minimise wavefront error (WFE) of return beam.
- **Test 2:** Collimated light from the interferometer is focused onto the image slicer by pre-optics (M1-4). The light passes through the IFU and the WFE is measured by a wavefront sensor (WFS) with a fixed focal length 100mm lens. This will verify the image slicer alignment.

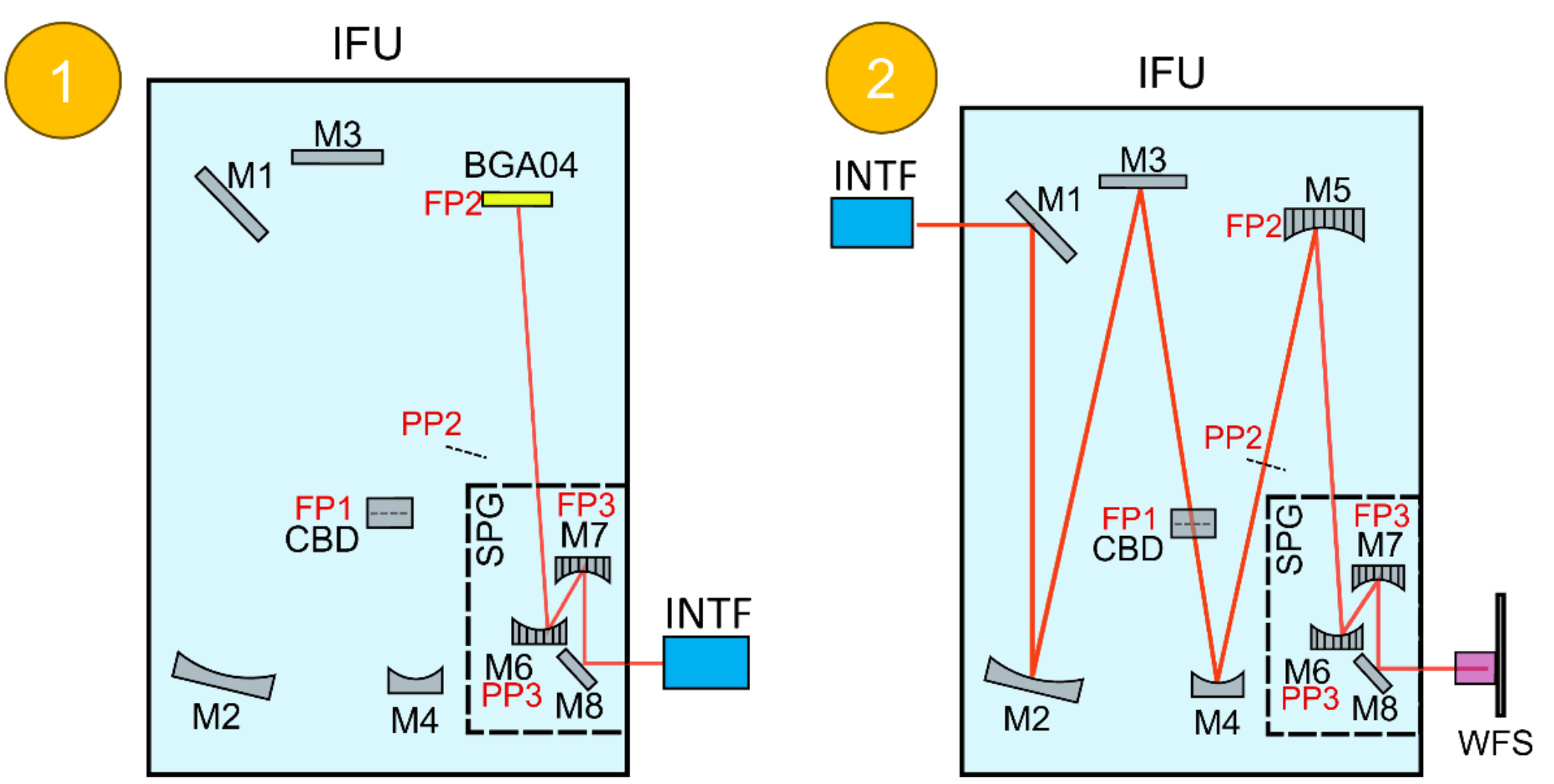

Figure 18. (left) IFU Test 1 and (right) IFU Test 2. FP1-3 and PP2-3 correspond to focal planes and pupil planes in the optical design. A coronographic beam dump (CBD) is placed at FP1.

To achieve the required alignment accuracy, the interferometer and wavefront sensor must be aligned to <75µm laterally relative to the IFU mirrors. This is done by measuring the position of the interferometer and wavefront sensor using spherically mounted retroreflector (SMR) targets or fixed calibration cones with a Vantage E6 laser tracker and a FARO Quantum E portable coordinate measuring machine. The SMR targets and calibration cones are mounted on metrology frames (referred to as “antlers”) to provide a longer baseline, giving higher angular accuracy (see Figure 19). Once the positions of the interferometer and wavefront sensor have been measured, each can be re-positioned using Symmetrie BORA hexapods. This methodology should position the interferometer and wavefront sensor with a maximum permissible error (MPE) of 68.4µm, giving confidence that the required alignment tolerance will be reached.

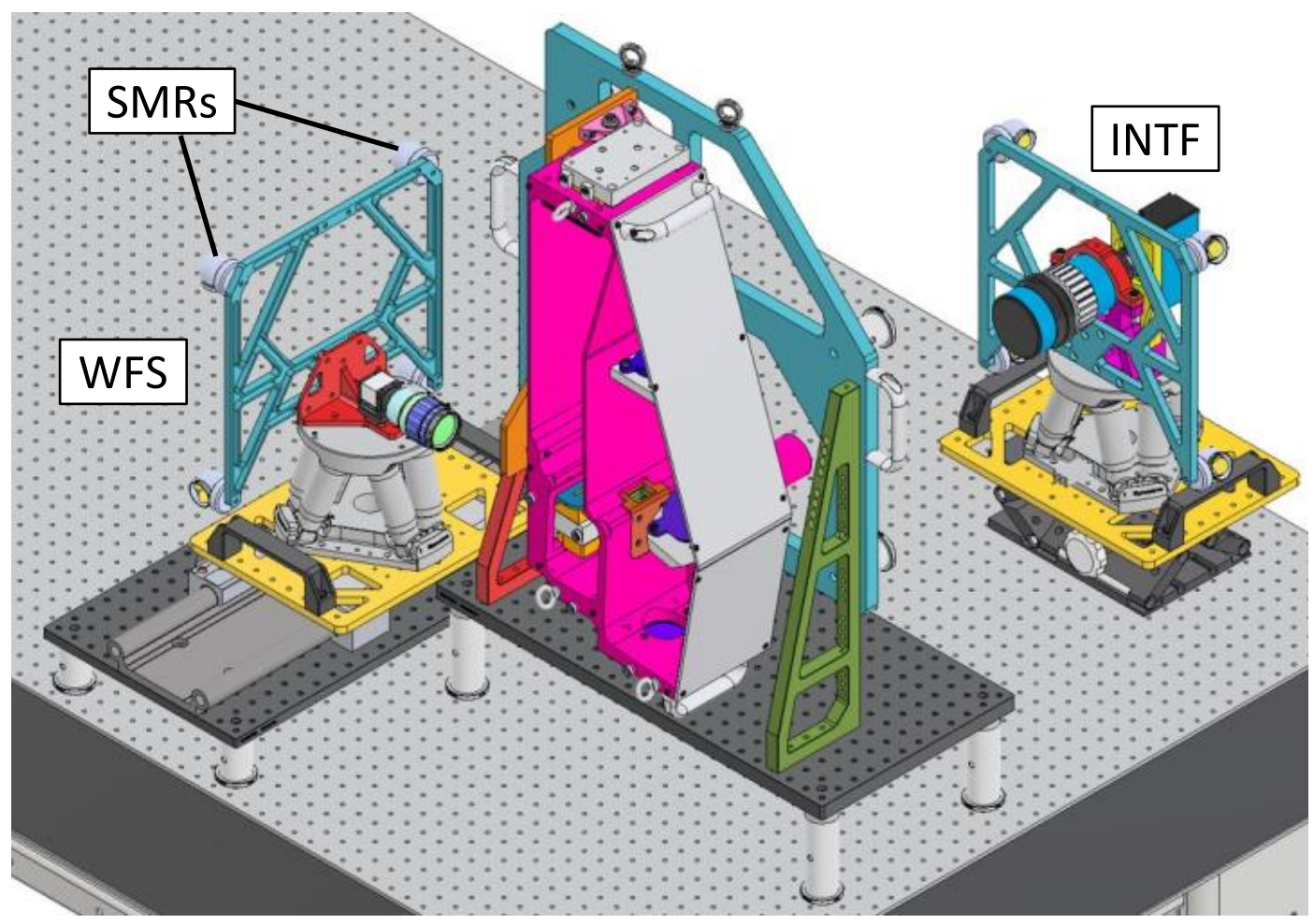

Figure 19. CAD showing setup for mechanical setup for IFU test 2. N.B., the "antlers" with mounted SMRs to allow for laser tracker metrology.

## 5 CONCLUSION

This proceedings presented the optical testing performed to verify the performance of the LMS M5 image slicer and an overview of the IFU optical tests that will be performed to align the optical components during the AIT process. The testing of the LMS M5 image slicer showed the performance summarised in Table 7. IFU REQ 13 and IFU REQ 14 were deemed compliant, while there was a non-compliance in IFU REQ 10 due to an error in the Cx and Cy positions of the slices. After analysis, the non-compliance was deemed acceptable and will be mitigated during AIT by precisely aligning the optical test equipment using both a laser tracker and portable coordinate measuring machine.

Table 7. Comparison of all the M5 Slicer requirements with the measured values during testing.

| Requirement | | | Measured Values |
|---|---|---|---|
| **IFU REQ 10.** | Cx | ±10µm | **±36.1µm** |
| | Cy | ±10µm | **±22.6µm** |
| | Cz | ±50µm | **±15.3µm** |
| **IFU REQ 13.** | Maximum Slope Error | <25µrad | **7.99µrad RMS** |
| | | | **99.90% within specification after removal of interferometer artifacts** |
| | | | **PVq/2 = 23.7µrad for q = 99.9%** |
| **IFU REQ 14.** | Surface form error over 5mm x 1mm sub-apertures | <10nm RMS | **Mean: 4.57nm RMS** |
| | | | **98.85% of all sub-apertures within 10nm requirement** |
| | | | **Max: 11.67nm RMS** |